# Unruh effect in a waveguide

Igor I. Smolyaninov

*Department of Electrical and Computer Engineering, University of Maryland, College Park, MD 20742, USA*

**ABSTRACT:** Here we demonstrate that photons launched into a specially designed metamaterial waveguide act as massive quasi-particles, which experience strong acceleration. Laser light propagating through such a waveguide may be used as a thermometer which would measure the Unruh temperature. Moreover, the metamaterial waveguide design may be approximated by a tapered optical waveguide.

Recent interest in optical metamaterials has been sparked by introduction of such concepts as a negative refractive index superlens [1-3], an electromagnetic invisibility cloak [4-6], etc., and realization that the transformation optics [4,5] may be used to design novel electromagnetic devices with unusual properties. The transformation optics approach is based on the conformal coordinate transformations, and thus on the behaviour of the electromagnetic field in curvilinear space-time metrics. Not surprisingly, the progress in metamaterial optics coincides with the introduction of optical analogs of black holes, and proposals to observe the Hawking radiation in tabletop fibre optical experiments [7].

In 1974 Hawking showed that black holes can evaporate by emission of thermal radiation [8]. A more general effect had been introduced a few years later by Unruh. He showed that for an accelerating observer vacuum should look like a bath of thermal radiation with temperature $T_U$ defined as

$$T_U = \frac{\hbar a}{2\pi kc} \qquad (1)$$

where $a$ is the acceleration of the observer [9]. The Hawking temperature may be obtained from this formula by substitution of the free fall acceleration near the black hole horizon into eq.(1). As has been shown by Unruh [9], the proper choice of the electromagnetic eigenmodes, and hence the proper choice of the electromagnetic vacuum in a reference frame moving with constant acceleration $a$ should be done based on the normal-mode solutions in the Rindler coordinates. The Rindler metric may be written as

$$ds^2 = \rho d\tau^2 - \frac{d\rho^2}{\rho} - dy^2 - dz^2 \qquad (2)$$

where $\tau=2at/c$, and $\rho=x^2/4$. For the sake of simplicity let us disregard the polarization degrees of freedom of the electromagnetic field, and consider the scalar theory. The normal mode solutions of the corresponding scalar wave equation can be found in ref. [9]. They can be written as

$$\psi_{\omega,k}^R = \frac{e^{-i\omega\tau}}{((2\pi)^3 \omega)^{1/2}} f(\rho) e^{-i(k_y y + k_z z)} \qquad (3)$$

where $f(\rho)$ satisfies the equation

$$\left(\rho\frac{d}{d\rho}\rho\frac{d}{d\rho} + \omega^2 - (k_y^2 + k_z^2)\rho\right)f(\rho) = 0 \qquad (4)$$

Unruh has shown that the coordinate transformation $t=2\rho^{1/2}sinh(\tau/2)$ and $x=2\rho^{1/2}cosh(\tau/2)$ converts the Rindler metric (eq.(2)) into the Minkowski one

$$ds^2 = dt^2 - dx^2 - dy^2 - dz^2 \qquad (5)$$

However, the vacuum state, which may be constructed based on the normal Minkowski eigenmodes $\psi^M_{\omega,k} \sim e^{-i(\omega t + kr)}$ looks like a thermal bath with temperature $T_U$. As a result, any detector, which moves through Minkowski vacuum with acceleration $a$ measures its temperature as $T_U$. While well-established theoretically, the Hawking effect and the Unruh effect are believed to be very difficult to observe in the experiment: an observer accelerating at $a$=9.8 m/s$^2$ should see "vacuum temperature" of only 4x10$^{-20}$ K. Over the past years quite a few proposals were introduced on how to emulate and/or observe these effects. The proposal by Philbin et al. [7] is making use of the optical waveguide arrangement in which right combination of Kerr nonlinearity and waveguide mode dispersion leads to an apparent event horizon for photons in a limited frequency range, and to potentially observable Hawking radiation (see also an almost identical proposal [10]). In this paper we consider a metamaterial waveguide arrangement which may lead to experimental observation of the more general Unruh effect. In principle, the metamaterial waveguide design may be approximated by a tapered optical waveguide. We demonstrate that the optical waveguide geometry may produce an accelerated motion of a collection of photons which behave as massive quasi-particles with internal degrees of freedom. The role of these internal degrees of freedom is played by the transverse modes of the metamaterial waveguide. An ensemble of these quasi-particles may be used as a test body to measure the Unruh temperature (eq.1).

We should point out that two kinds of temperatures may be introduced for this system: the temperature of the lateral 1D motion of the photon quasi-particles in the co-moving reference frame, and the temperature of the internal degrees of freedom, which corresponds to the mode coupling in the waveguide (this temperature is similar to the spin temperature of the nuclear spin system). Both temperatures are supposed to be equal to the Unruh temperature. Thus, mode coupling in an adiabatically tapered waveguide may be considered as a manifestation of the Unruh effect. An example of such a mode-coupling experiment in a tapered waveguide is presented. We should also mention that the waveguide geometry emulates theories with a compactified extra dimension, such as the Kaluza-Klein theories (see for example ref. [11]), in which "normal" particles are treated as the modes of the electromagnetic field guided by a thin extra-dimensional waveguide. Our result appears to be completely natural and expected in view of this analogy.

Let us consider an optical waveguide shown in Fig.1(a). Let us assume that all walls of the waveguide are made of an ideal metal. The waveguide is filled with vacuum. The waveguide has constant height $d$ and width $b$ in the $z$- and $y$- directions, respectively. Let us assume that this waveguide is either immersed in a weak constant gravitational field, or is subjected to accelerated motion with constant acceleration (these are locally equivalent situations). Since the gravitational field is static, this geometry may be represented by a metamaterial waveguide in which the refractive index changes as a function of $x$-coordinate. The equations of electrodynamics in the presence of static gravitational field look exactly like Maxwell equations in the medium in which $\varepsilon=\mu=g_{00}^{-1/2}$ (see ref. [12]). Since the gravitational field is weak, $g_{00}=1+2\phi/c^2$, where $\phi=-gx$ is the gravitational potential [12]. Thus, the waveguide subjected to gravitational field (or experiencing constant acceleration) may be represented by a metamaterial waveguide in which $\varepsilon=\mu$ and both $\varepsilon$ and $\mu$ have a gradient in the $x$-direction, so that $n=(\varepsilon\mu)^{1/2}=1+gx/c^2$. In other words, the "optical dimensions" of the metamaterial waveguide ($n(x)d$ and $n(x)b$) change with the $x$-coordinate. If the optical terminology is used, such waveguide is called a tapered waveguide. On the other hand, in the presence of the refractive index gradient the effective acceleration of the waveguide may be obtained as

$$a = c^2 \frac{dn}{dx} \qquad (6)$$

If the refractive index changes fast, the effective acceleration of the waveguide is rather large: $\Delta n \sim 0.3$ over $\Delta x \sim \lambda \sim 200$nm produces effective acceleration of the order of $a \sim 10^{23}$ m/s$^2$, which corresponds to the Unruh temperature $T_U \sim 540$ K (see eq.(1)). We are going to show that laser-emitted coherent photons which move through such a waveguide may play the role of a temperature detector capable of measuring the Unruh temperature.

For the sake of simplicity, let us now consider the tapered waveguide geometry shown in Fig.1(b), which is easier to realize in the experiment, and once again consider the scalar theory. These assumptions will allow us to illustrate the basic physics ideas behind the proposal to observe the Unruh effect in a metamaterial waveguide. Tapering the waveguide achieves the same qualitative result as the refractive index gradient. However, the exact equivalence with the behavior of Maxwell equations in a gravitational field will be lost.

Let us assume that all the walls of the waveguide are made of an ideal metal and the waveguide is filled with vacuum. The waveguide is very thin and has a constant height $d$ in the $z$-direction: $d=const\sim\lambda/2$, where $\lambda$ is the wavelength of laser light used in the experiment. The dispersion law of photons propagating inside this waveguide looks like a dispersion law of a massive quasi-particle with an effective mass $m^*=\pi\hbar/(cd)$:

$$k_x^2 + k_y^2 + k_z^2 = k_x^2 + k_y^2 + \frac{\pi^2}{d^2} = \frac{\omega^2}{c^2} \qquad (7)$$

Various transverse modes of this waveguide are described by the mode number $n_y$ as $k_y = n_y \pi / b$, where $b$ is the width of the waveguide in the $y$-direction, which is assumed to be large compared to $\lambda$. The mode number $n_y$ may be considered as an internal degree of freedom of the photon quasi-particle, which is similar to spin. We will assume that originally the photons are launched by laser light into a state with some particular $n_y = n_0$, and the rest of the transverse modes are not excited. Since the laser light is coherent, all the quasi-particles have the same initial velocity in the $x$-direction, and hence the initial temperature of the one-dimensional gas of these quasi-particles is close to zero in the co-moving reference frame. At the same time, since for all quasi-particles $n_y = n_0$, the initial temperature corresponding to the internal degrees of freedom (similar to the spin temperature in the nuclear spin system) is close to zero too.

In order to produce an accelerated motion of these quasi-particles the width of the waveguide $b(x)$ in the $y$-direction may be changed adiabatically as a function of $x$. Under such an adiabatic change the photons are going to stay in the same lateral mode $n_y = n_0$ [13] (see the discussion below) so we can write for $k_x$ an approximate expression

$$k_x^2 = \frac{\omega^2}{c^2} - \frac{\pi^2}{d^2} - \frac{\pi^2 n_0^2}{b^2(x)} \qquad (8)$$

Since $d \sim \lambda/2$, the quasi-particle is non-relativistic and its acceleration $a$ may be calculated as

$$a = v_g \frac{dv_g}{dx} = \frac{\pi^2 n_0^2 c^4}{\omega^2 b^3}\left(\frac{db}{dx}\right) \qquad (9)$$

where $v_g = d\omega/dk_x$ is the group velocity of the quasi-particle. An observer in the reference frame co-moving with the quasi-particle with the same acceleration $a$ should perceive vacuum as a thermal bath with the Unruh temperature

$$kT_U = \frac{\hbar \pi n_0^2 c^3}{2\omega^2 b^3}\left(\frac{db}{dx}\right) \qquad (10)$$

The walls of the waveguide will have the same temperature $T_U$ in the co-moving frame. The internal degrees of freedom of the quasi-particle are supposed to thermalize at the same temperature $T_U$. In the accelerated reference frame, which is co-moving with a given photon quasi-particle, the energy level splitting between the internal degrees of freedom is equal to

$$\Delta\omega = \frac{\pi^2 c^2 n_0 \Delta n_y}{b^2 \omega} \qquad (11)$$

(see eq.(8)). Thus, there should be a mode coupling described by the Boltzmann's factor

$$C_{\Delta n_y} = e^{-\frac{\hbar \Delta \omega}{kT_U}} = \exp\left(-\frac{2\pi\omega b \frac{\Delta n_y}{n_0}}{c\frac{db}{dx}}\right) \qquad (12)$$

It is important to notice that the thermal distribution of mode numbers $\Delta n_y$ described by equation (12) is accompanied by the corresponding thermal distribution of $\Delta k_x$ in the co-moving reference frame (due to energy conservation: note that a passive waveguide cannot create quanta on its own). While in the laboratory reference frame the frequency of all photons remains the same, and the number of photons is conserved, in the co-moving accelerating reference frame the observed distribution of $\Delta k_x$ would broaden due to Doppler effect, since photons with different $n_y$ have different $x$-components of their speed. Upon accelerated propagation through a tapered waveguide the one-dimensional gas of photon quasi-particles acquires the Unruh temperature as defined by equation (10) for both the internal degree of freedom described by $\Delta n_y$, and for the lateral movement along the waveguide described by $\Delta k_x$. Thus, we have demonstrated that the Unruh effect should be partially responsible for the mode coupling in an adiabatically tapered waveguide. In order to achieve complete analogy, a metamaterial waveguide must be used in which $\varepsilon=\mu$ and both $\varepsilon$ and $\mu$ have a gradient in the x-direction, so that $n=(\varepsilon\mu)^{1/2}=1+ax/c^2$. In addition, the effect described by eq.(12) should be distinguishable from the usual classic mode coupling in a waveguide. How difficult is to distinguish these effects?

In order to answer this question we are going to approximate the portion of the adiabatically tapered waveguide in Fig.1(b) as a waveguide step shown in Fig.1(c) (this approximation also works for a waveguide shown in Fig.1(a), in the latter case $b_0$ and $b_1$ would be the "optical widths" of the respective portions of the waveguide defined as $n(x)b$ ). In the simplified scalar consideration the lateral modes in the respective sections of the step waveguide are described by $\phi^0_p=exp(i\pi py/b_0)$ and $\phi^1_q=exp(i\pi qy/b_1)$, respectively. The mode coupling coefficients are obtained as

$$C^{1/2}_{pq} = \frac{1}{b_1}\int_{-b_1/2}^{b_1/2} \phi^0_p \phi^{1*}_q dy = \frac{\sin\left(\pi b_1\left(\frac{p}{b_0}-\frac{q}{b_1}\right)\right)}{\pi b_1\left(\frac{p}{b_0}-\frac{q}{b_1}\right)} \qquad (13)$$

Introducing $\Delta n_y=(q-p)$, $p=n_0$, and $\Delta b=b_1-b_0$ we can rewrite eq.(13) as

$$C_{\Delta n_y} = \frac{\sin^2\left(\pi n_0 \frac{\Delta b}{b}\right)}{\pi^2\left(\Delta n_y - n_0 \frac{\Delta b}{b}\right)^2} \qquad (14)$$

which indicates that photons are indeed most probably staying in the same lateral mode $n_y=n_0$. However, in the scalar theory the quantum Unruh effect described by eq.(12) is considerably weaker than the classic mode coupling described by eq.(14). On the other hand, if the polarization degree of freedom of the photon quasi-particles is taken into account, the classic expression for the mode coupling is exactly zero for the orthogonal polarization states in the waveguide. This is not the case for the Unruh effect, which is thermodynamic in nature. Thermal fluctuations of the waveguide walls, which have the Unruh temperature in the co-moving accelerating reference frame will lead to mode coupling described by eq.(12) independent of the polarization state. Thus, unambiguous observation of the Unruh effect in a metamaterial waveguide would need to be performed in a highly symmetric waveguide at low temperatures, so that there would be no cross-talk between different polarization states in the waveguide due to such classic effects as waveguide imperfections and thermal fluctuations of the waveguide shape in the laboratory reference frame. On the other hand, in the "fast taper" case shown in Fig.2 (in this experiment $db/dx\sim0.25$ at $\lambda=632$ nm) the effects described by eq.(12) and (14) have the same order of magnitude. The mode coupling effect is large and easily detectable. In the experiment presented in Fig.2 a single mode fibre is adiabatically tapered to sub-micrometer dimensions, so that the core and cladding of the original fibre become the core of a tapered multimode optical waveguide shown in Fig.2(a). Laser light originally propagating through the single mode portion of the fibre, couples to large number of transverse modes upon propagation through this multimode tapered waveguide. This is evident from the multiple rings observed in the mode distribution measured at the apex of the tapered fibre using a near-field scanning optical microscope. Using the language of the Unruh effect, we may say that the effective temperature, which corresponds to the excitation of the internal degrees of freedom of the photon quasi-particles is comparable to the mode spacing: $kT_U \sim \hbar \Delta \omega$. Indeed, the Unruh temperature perceived by the photon quasi-particles propagating through the tapered fibre in Fig.2(a) may be estimated from the geometry of the taper and equation (10) as a few hundreds of Kelvin. The mode spacing estimated from eq.(11) has the same order of magnitude. As a result, the mode coupling is very strong.

**Figure Captions**

Fig.1 (a) Optical waveguide which is subjected to accelerated motion. The effective refractive index gradient is shown by halftones. (b) Geometry of the adiabatically tapered waveguide. (c) Approximation of the tapered waveguide from (a) or (b) as a step waveguide.

Fig.2 (a) Electron microscope image of an adiabatically tapered optical fiber. (b) Mode distribution measured at the apex of the fiber using scanning near-field optical microscope at 632 nm laser wavelength.

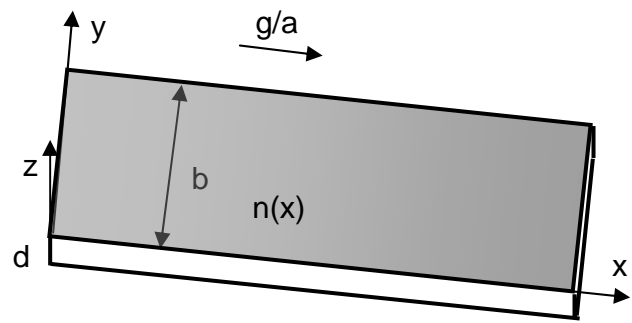

(a)

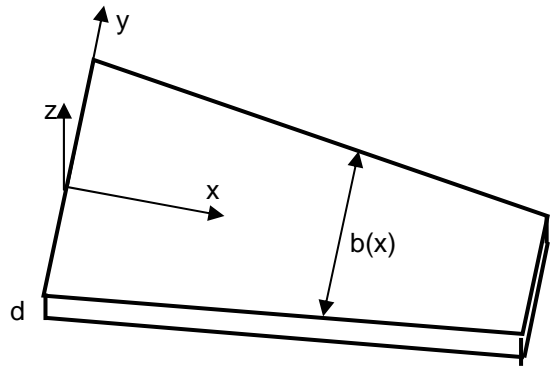

(b)

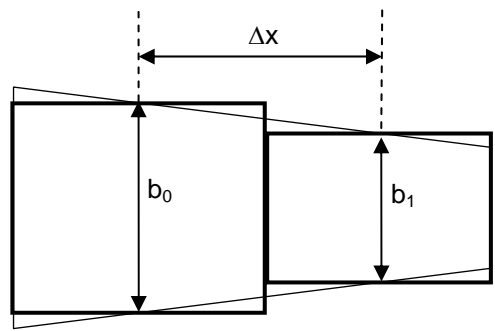

(c)

Fig.1

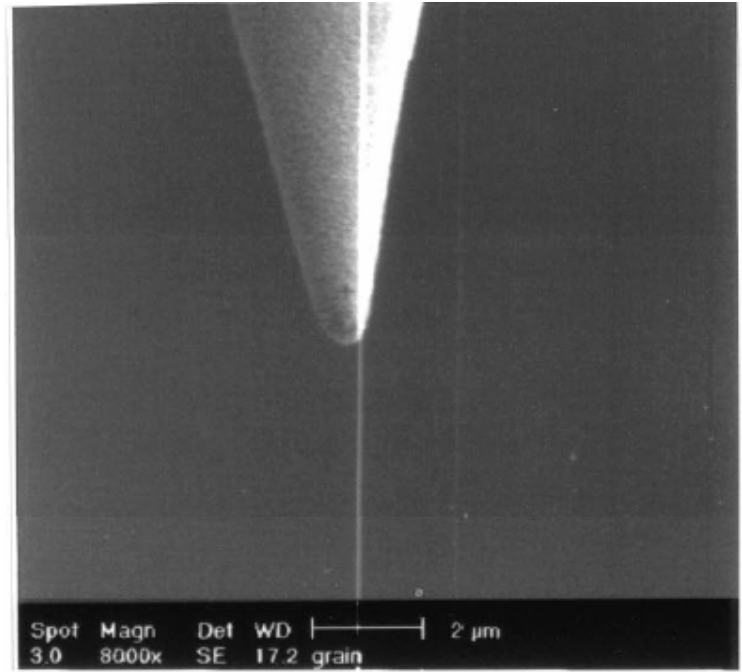

( a )

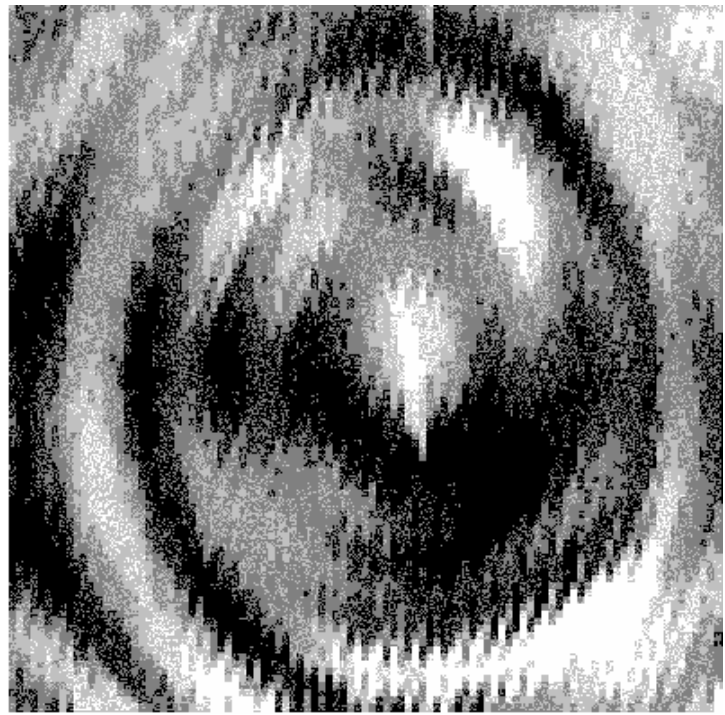

( b )

Fig.2